# Weighted Simultaneous Algebra Reconstruction Technique (wSART) for Additive Light Field Synthesis

*Chen Gao\*, Linqi Dong\*, Liang Xu\*, Xu Liu\*, Haifeng Li\**
*\*State Key Lab. of Modern Optical Instrumentation, Zhejiang University, Hangzhou, China*

## Abstract
*We apply an iterative weighting scheme for additive light field synthesis. Unlike previous work optimizing additive light field evenly over viewpoints, we constrain the optimization to deliver a reconstructed light field of high image quality for viewpoints of large weight.*

## 1. Introduction
Additive light field displays use multiple transparent and self-luminous display layers to provide vergence cues in the autostereoscopic 3D display or accommodation cues in near-eye displays [1-3]. Compared with multiplicative counterparts suffering diffraction and brightness loss [4, 5], the additive light field display is relatively free from these problems, thus suitable for augmented reality application. However, since the reconstructed error could only accumulate on each layer, the additive light field displays fail to provide a large field of view (FOV) by time multiplexing [6].

Previous work on additive light field displays [1-3, 7] uses a trust-region algorithm or simultaneous algebra reconstruction technique (SART) to solve display patterns evenly for all viewpoints. These optimization algorithms can not handle too many different parallaxes of target light fields with a large FOV, resulting in reconstructed light fields of poor image quality. Inspired by recent work on discrete tomography [8] and performance improvement for multiplicative light field displays [9], we propose a weighted simultaneous algebraic reconstruction technique (wSART) scheme for wide FOV additive light field reconstruction.

## 2. Display System and Principle
As Figure 1 shows, the display system comprises dual-layer transparent organic light-emitting diode (OLED) panels, a diffuser inserting between dual-layers (not shown), and a tracking device. The reconstructed light field $L$ emitted by the self-luminous dual-layer is expressed as:

$$L = I_1 + I_2 \quad (1)$$

where $I_1$ and $I_2$ are light intensity of display patterns of front and rear transparent OLED panels, respectively. Rewrite Equation 1 in the form of matrix multiplication as Equation 2:

$$L_i = \sum_{j=1}^{N} P_{i,j} I_j \quad (2)$$

where $L_i$ is the $i^{th}$ viewpoint image, the $P_{i,j}$ denotes the $(i, j)$ entry of projection matrix $P$ that stores the mapping between light field rays and layer pixels. For this kind of additive light field display, we are compelled to solve the following problem:

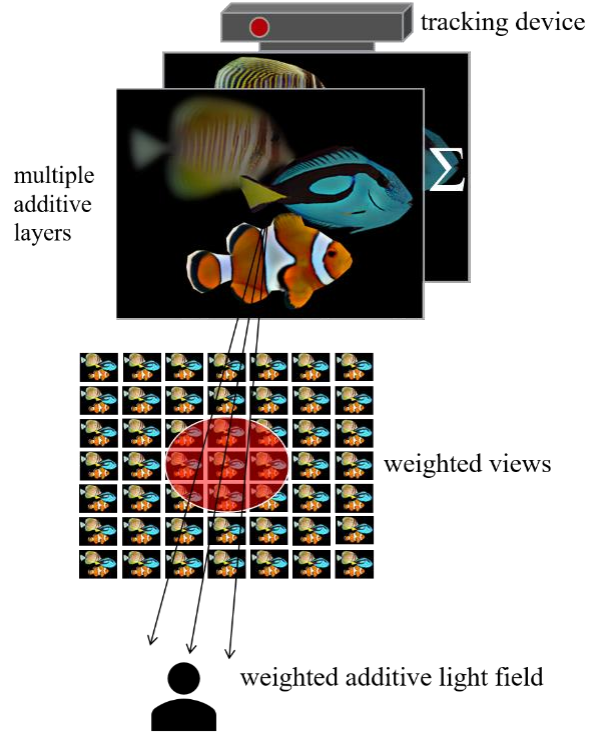

**Figure 1.** Overview of the wide FOV additive light field display using wSART and a tracking device. Two transparent OLED panels reconstruct a light field in an additive manner. A view-position-dependent weight distribution function is generated based on the eyes' positions acquired by the -tracking device. When standing in front of the display system, all viewers could perceive binocular parallax of high image quality.

$$\arg\min \|L_t - PI\| \quad (3)$$

where $L_t$ is the target light field. The tracking device acquires viewers' eye positions and generates corresponding weight distribution functions for the factorization algorithm.

## 3. Viewing-position-dependent weight optimization
Figure 2 shows the layout of the projection matrix $P$. The matrix's gray elements in the $(i, j)$ entry represent pixels on the $j^{th}$ layer hit by the $i^{th}$ viewpoint's emitting rays. To solve Equation 3 using the SART scheme, whose iteration rule is illustrated as Equation 4, which means every row of the projection matrix $P$ has the same weight. When the parallaxes between viewpoint images become large as the target light field's FOV increases, the SART algorithm is hard to deal with too much dissimilarity, and the reconstruction quality falls. To enable additive light field display of a wide FOV while delivering high image quality to specific views, we apply a wSART scheme with a tracking device, whose iteration rule is written as Equation 5, where $I_j^{(k)}$ is the $j^{th}$ display layer pattern from the $k^{th}$ iteration and $w_i \in [0,1]$ is the weight for the $i^{th}$ viewpoint.

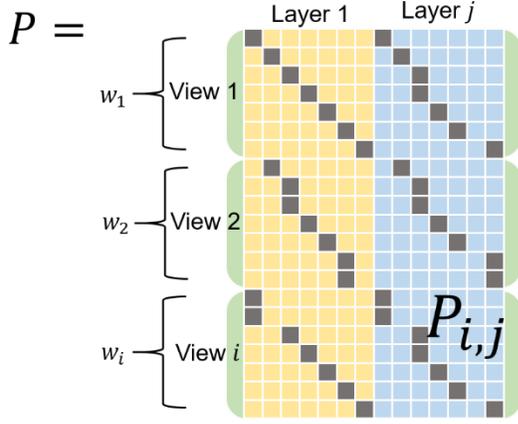

**Figure 2.** The layout of the projection matrix. The matrix's gray elements in the *(i, j)* entry represent pixels on the $j^{th}$ layer hit by $i^{th}$ viewpoint's rays.

$$I_j^{(k+1)} = I_j^{(k)} + \frac{\sum_i P_{i,j}^T \left( \frac{L_t - \sum_j P_{i,j} I_j^{(k)}}{\sum_j P_{i,j}} \right)}{\sum_i P_{i,j}} \qquad (4)$$

$$I_j^{(k+1)} = I_j^{(k)} + \frac{\sum_i P_{i,j}^T w_i \left( \frac{L_t - \sum_j P_{i,j} I_j^{(k)}}{\sum_j P_{i,j}} \right)}{\sum_i P_{i,j} w_i} \qquad (5)$$

Let $(x_i, y_i)$ be the horizontal and vertical coordinates of the $i^{th}$ viewpoint located on the viewing plane. We list three schemes of weight distribution functions as below (Figure 3).

*(a) Even weight function:*

$$w_i = 1 \qquad (6)$$

When substituting even weight function into wSART, Equation 5 reduces exactly to the standard SART (Equation 4).

*(b) Binary weight function:*

$$w_i = \begin{cases} 1 & \text{for } x_{lb} \leq x_i \leq x_{ub} \text{ and } y_{lb} \leq y_i \leq y_{ub} \\ 0 & \text{otherwise,} \end{cases} \qquad (7)$$

where $(x_{lb}, x_{ub})$ and $(y_{lb}, y_{ub})$ are lower and upper bounds of viewpoint positions in horizontal and vertical directions.

*(c) Gaussian weight function.*

The Gaussian weight function can be decomposed in two perpendicular directions:

$$w_i = \exp\left( \frac{-(x_0 - x_i)^2}{2\sigma_x^2} \right) \exp\left( \frac{-(y_0 - y_i)^2}{2\sigma_y^2} \right) \qquad (8)$$

where $(x_0, y_0)$ is the center of the viewer's head. $\sigma_x$ and $\sigma_y$ are standard deviations that determine the shape of the Gaussian weight function in x and y directions, respectively.

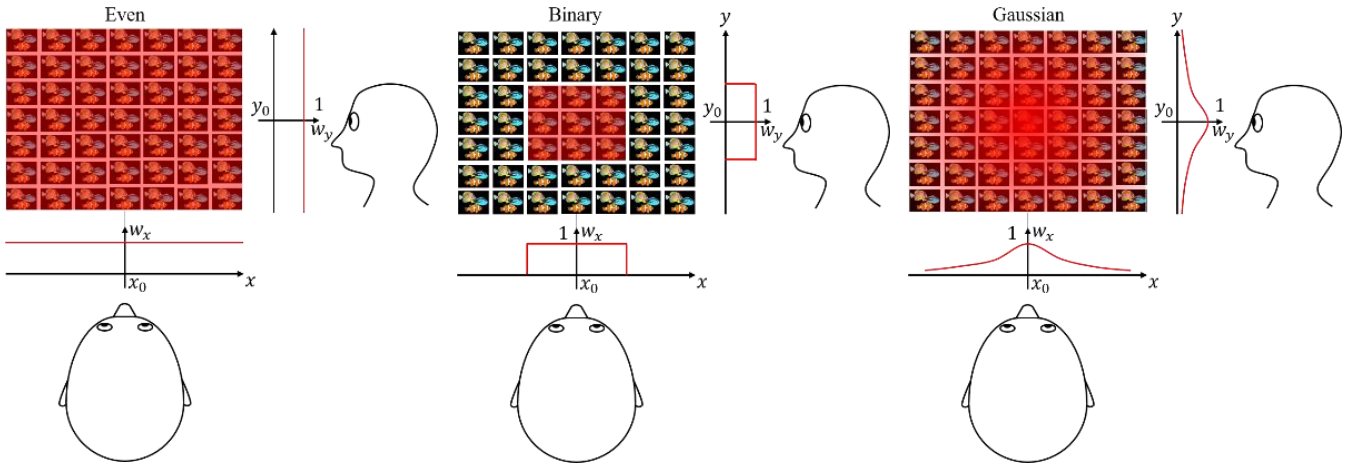

**Figure 3.** The schematic diagram of three expressions of weight distribution functions. (left) The Even weight function for basic additive light field synthesis. (middle) The Binary weight function for the basic additive light field synthesis in tracking. (right) The Gaussian weight function for weighted additive light field synthesis.

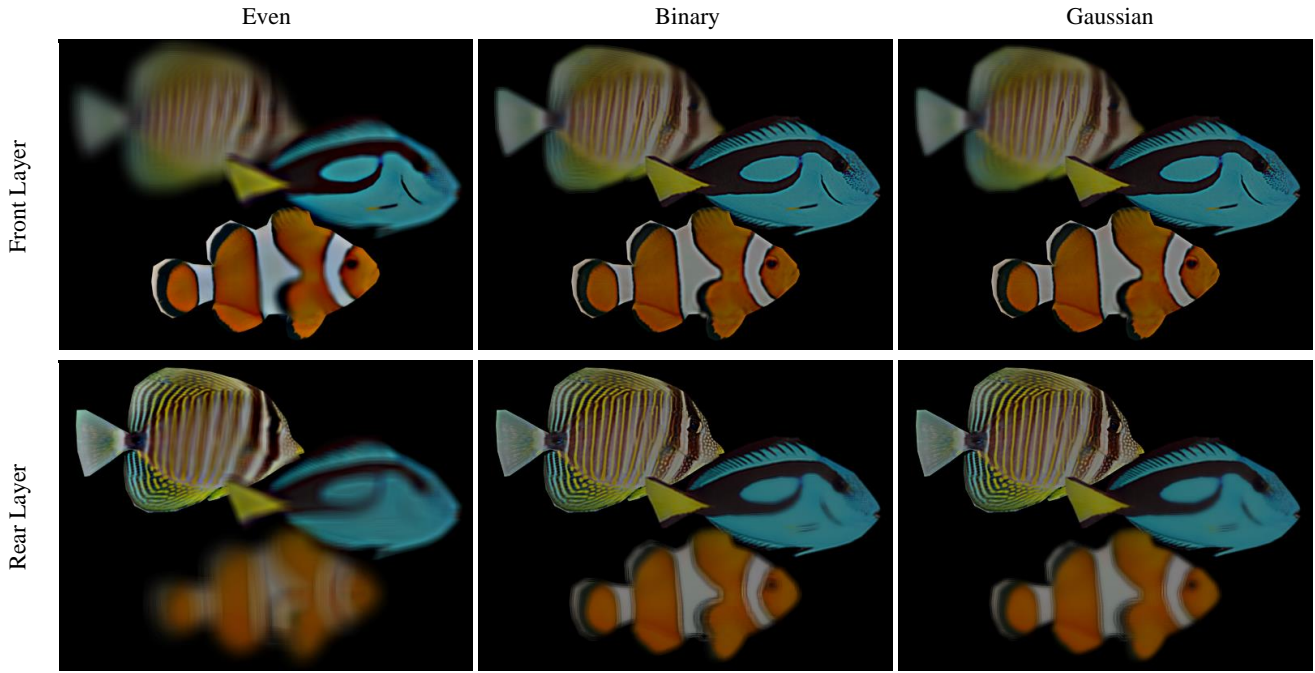

**Figure 4.** Factorized dual-layer display patterns of three weight distribution functions.

## 4. Implementation and results

***Hardware:*** We built one additive light field tabletop display to demonstrate the validity of the wSART scheme. The prototype uses dual-layer transparent OLED panels of LG Display LW550JUL. The moiré pattern (see the upper part of the display area of the prototype in Figure 5) is eliminated by inserting a 4mm thick circular light shaping diffuser on PET (polyethylene terephthalate) from Luminit with a diffusing angle of 5° between dual-layers. The results presented below were obtained on an Intel Core i7-4790 PC with 16 GB of RAM and an NVIDIA GeForce GTX 2080Ti graphics card. Experimental results are captured with a digital single-lens reflex (DSLR) camera (Canon EOS 5D Mark II) with a resolution of 4320×3240 in the dark.

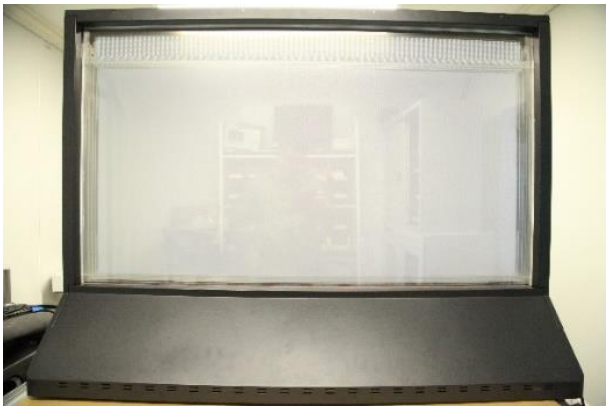

**Figure 5.** Photograph of the additive light field display.

***Software:*** We implemented a MATLAB-based offline algorithm for the wSART scheme. The target light fields consist of 9×9 views that capture parallax images of a 3D tropical fish scene within a FOV of 15°×15°. The spatial resolution of each parallax image is 512×384 pixels. For all three weight distribution functions with 5 iterations, it takes 1.9 seconds to solve the display patterns for each color channel. Figure 4 shows factorized dual-layer display patterns using the three weight distribution functions above. The viewer's head center is at the central target view. For the binary weight function, the lower and upper bounds are the $4^{th}$ and $6^{th}$ target view's positions in horizontal and vertical directions. For the gaussian weight function, the standard deviations along the x and y directions are set to 100. Alignment and display of the dual-layer patterns on the transparent OLEDs is performed using an OpenGL program.

***Result:*** Figure 6 shows a simulated and experimental reconstructed light field of the central viewpoint of three weight distribution functions. For experimental results, we set the lens' focal length (EF 24-105mm f/4L IS USM) to 58 mm, the F-number to 4. These photographs were captured with a shutter speed of 1/60 second and an ISO speed of 1000. The simulation's peak signal-to-noise ratio (PSNR) values are marked with red letters in the upper right corners. The red and green regions where their enlargements are shown on the bottom side of the figure, located at the fin part of clownfish fish, present significant image quality improvements with binary and gaussian weights. The result shows that the Gaussian weight distribution functions reached the highest PSNR of value 31.54 dB.

## 5. Conclusion

We have applied an adaptive weighting scheme to achieve high quality reconstructed light field for specific viewpoints. And this method is verified by both simulation and experiment. While the current reconstructed scene is static, we are working on a graphics processing unit (GPU) acceleration program for the wSART scheme with a tracking device to achieve real-time rendering of a wide FOV.

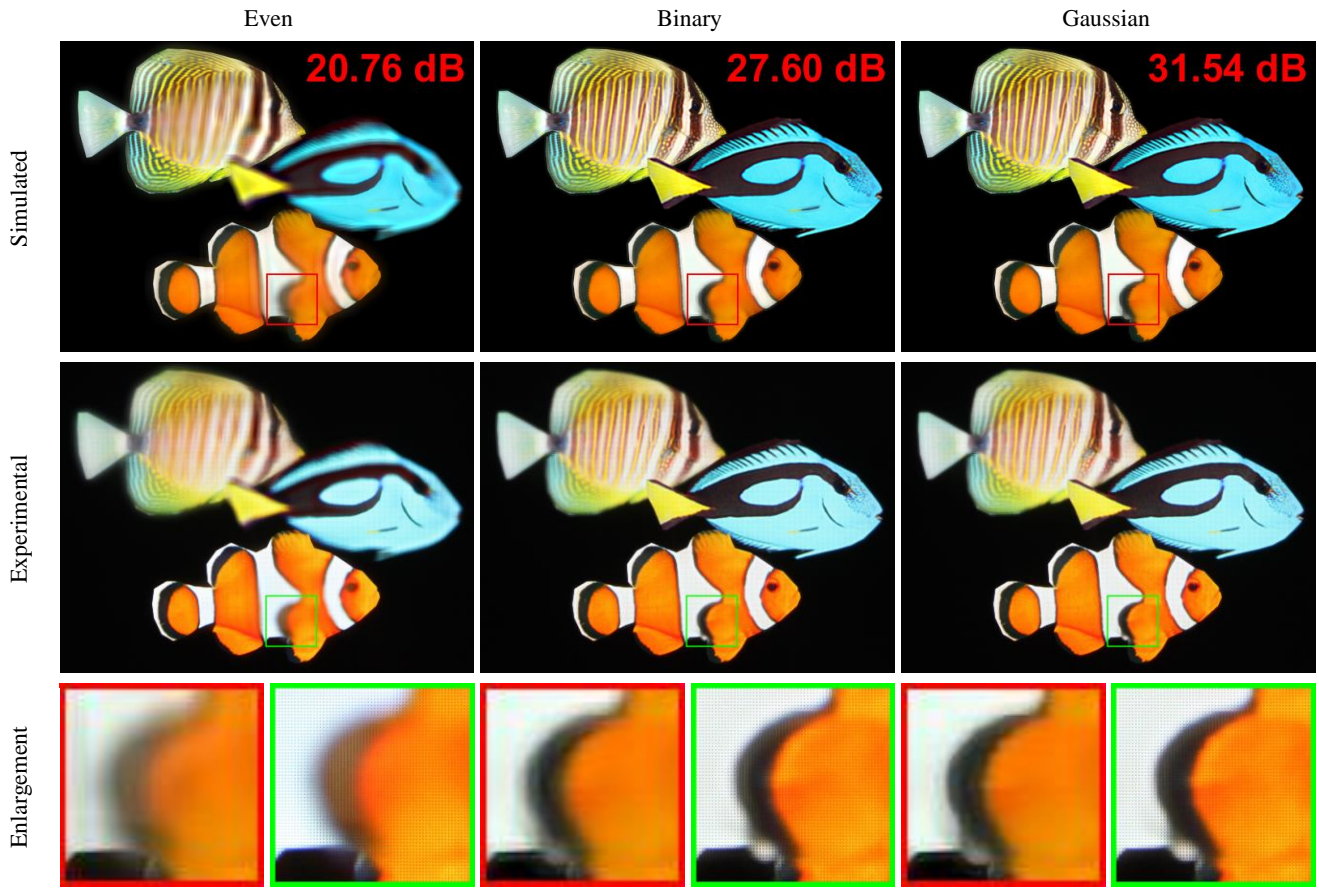

**Figure 6.** Simulated and experimental reconstructed light field of the central viewpoint of three weight distribution functions. The PSNR values for simulation are marked with red letters in the upper right corners. The red and green regions where their enlargements are shown on the bottom side of the figure, located at the fin part of clownfish fish, present significant image quality improvements with binary and gaussian weights.

## 6. Acknowledgments

This work is supported by the National Key Research and Development Project of China (2017YFB1002900).